\documentclass[aps,prd,showpacs,amsmath,amssymb,superscriptaddress,floatfix,nofootinbib,10pt]{revtex4}
\usepackage{times}
\usepackage{amssymb,amsbsy,amsmath,amsfonts}
\usepackage{graphicx}
\usepackage{float}
\usepackage{color}
\usepackage{morefloats}
\usepackage{rotating}
\usepackage{srcltx}
\usepackage{slashed}
\usepackage{multirow}
\usepackage{verbatim}
\usepackage{hyperref}
\usepackage{tabularx}
\usepackage{bm}
\allowdisplaybreaks[4]

\usepackage{bbding}
\usepackage{threeparttable}

\newcommand{\PreserveBackslash}[1]{\let\temp=\\#1\let\\=\temp}
\newcolumntype{C}[1]{>{\PreserveBackslash\centering}p{#1}}
\newcolumntype{R}[1]{>{\PreserveBackslash\raggedleft}p{#1}}
\newcolumntype{L}[1]{>{\PreserveBackslash\raggedright}p{#1}}

\begin{document}

\title{$^3S_1-{}^3D_1$ coupled channel $\Lambda_c N$ interactions: chiral effective field theory vs. lattice QCD}

\author{Jing Song}
	\affiliation{School of space and environment, Beihang University, Beijing, 102206, China}
\affiliation{School of Physics, Beihang University, Beijing, 102206, China}

\author{Yang Xiao}
	\affiliation{School of space and environment, Beihang University, Beijing, 102206, China}
\affiliation{School of Physics, Beihang University, Beijing, 102206, China}
\affiliation{Universit\'e Paris-Saclay, CNRS/IN2P3, IJCLab, Orsay, 91405, France}

\author{Zhi-Wei Liu}
\affiliation{School of Physics, Beihang University, Beijing, 102206, China}

\author{Kai-Wen Li}
\email[E-mail me at: ]{kaiwen.li@buaa.edu.cn}
\affiliation{Medical Management Department, CAS Ion Medical Technology Co., Ltd., Beijing 100190, China}
\affiliation{Beijing Advanced Innovation Center for Big Data-Based Precision Medicine, School of Medicine and Engineering, Beihang University, Key Laboratory of Big Data-Based Precision Medicine (Beihang University), Ministry of Industry and Information Technology, Beijing, 100191, China}
\affiliation{School of Physics, Beihang University, Beijing, 102206, China}

\author{Li-Sheng Geng}
\email[E-mail me at: ]{lisheng.geng@buaa.edu.cn}
\affiliation{School of Physics, Beihang University, Beijing, 102206, China}
\affiliation{Beijing Key Laboratory of Advanced Nuclear Materials and Physics, Beihang University, Beijing, 102206, China}
\affiliation{School of Physics and Microelectronics, Zhengzhou University, Zhengzhou, Henan, 450001, China}

\begin{abstract}
We study the lattice QCD  $\Lambda_c N$ phase shifts for the $^3S_1-{}^3D_1$ coupled channel using both the leading order covariant chiral effective theory and the next-to-leading order non-relativistic chiral effective field theory. We show that although it is possible to describe simultaneously the $^3S_1$ and $^3D_1$ phase shifts  and the inelasticity $\eta_1$, the fitted energy range is pretty small, only up to $E_\mathrm{c.m.}=5$ MeV. This raises concerns regarding the consistency between leading/next-to-leading order  chiral effective field theory and  the lattice QCD simulations.

\end{abstract}


\date{\today}

\maketitle
\section{Introduction}
The $\Lambda_c$ baryon as the lightest charm baryon has attracted a lot of attention, which may exist in finite nuclei to form  $\Lambda_c$ hypernuclei. The HAL QCD Collaboration performed the first lattice QCD simulations of the $\Lambda_c N$ and $\Sigma_cN$ interactions for unphysical light quark masses ($m_\pi=410,~570,~700$ MeV)~\cite{Miyamoto:2017tjs}, which provided vital information on the interaction between a nucleon and a charmed baryon $\Lambda_c$ or $\Sigma_c$. Employing these lattice QCD results, extrapolations to the physical point have been performed using either the non-relativistic chiral effective field theory (ChEFT) at next-to-leading order (NLO)~\cite{Haidenbauer:2017dua} or the covariant ChEFT at leading order (LO)~\cite{Song:2020isu}. In the  covariant ChEFT, Lorentz covariance is maintained by employing the covariant chiral Lagrangians, the full form of Dirac spinors, and the relativistic scattering equation (the Kadyshevsky equation). It has been shown that the covariant ChEFT approach can provide  reasonable  descriptions of octet baryon-octet baryon interactions already at LO, including all the systems from strangeness $S=0$ to $S=-4$, at least in the low energy region~\cite{Ren:2016jna,Li:2016mln,Ren:2017yvw,Song:2018qqm,Ren:2018xxd,Li:2018tbt,Wang:2020myr,Bai:2020yml,Liu:2020uxi,Song:2021yab,Liu:2022nec}.~\footnote{The next-to-next-to-leading order relativistic chiral nucleon-nucleon interaction is shown to be able to describe the neutron-proton scattering phaseshifts up to $T_\mathrm{lab.}=200$ MeV as well as the next-to-next-to-next-to-leading order non-relativistic chiral nucleon-nucleon interactions.~\cite{Lu:2021gsb}.} A recent study~\cite{Bai:2020yml} showed that one could reproduce both the physical $^1S_0$ and  $^3S_1-{}^3D_1$   and the lattice QCD nucleon-nucleon partial wave phase shifts fairly well. In particular, for the physical nucleon-nucleon phase shifts and lattice QCD data at  $m_\pi=469$ MeV, if one only fits to the $^3S_1$ phase shifts, the  $^3D_1$ phase shifts and inelasticity $\eta_1$ can be predicted and vice versa, as shown in Ref.~\cite{Bai:2021uim}. It implies that indeed the correlations induced by the imposed  constraint of covariance in the covariant chiral potentials  is reasonable.

In our previous study of the $\Lambda_c N$ interaction in the covariant ChEFT~\cite{Song:2020isu}, the low energy constants (LECs) were determined by fitting to the lattice QCD data from the HAL QCD Collaboration, where the $S$-wave phase shifts up to $E_{\textrm{c.m.}}=30$ MeV for  $m_\pi=410$ MeV and $570$ MeV were considered. The results showed that the covariant ChEFT can describe the lattice QCD data fairly well at low energies. In addition, the phase shifts of the $\Lambda_c N$ $^3D_1$ partial wave and the inelasticity $\eta_1$, as well as their physical counterparts were predicted. 

In a recent study~\cite{Haidenbauer:2021tlk}, it was shown that the predicted $^3D_1$ phase shifts by the NLO non-relativistic ChEFT are in agreement with the lattice QCD data of Ref.~\cite{Miyamoto:2019mfk} at higher energies, but not those of Ref.~\cite{Song:2020isu}. A closer examination of the lattice QCD data revealed, however, that although at higher energies, the predictions of Ref.~\cite{Song:2020isu} do not agree with the lattice QCD data, but at low energies close to threshold, they do agree, both for the $^3D_1$ phase shifts and the inelasticity, at least qualitatively. On the other hand, the predictions of the NLO non-relativistic ChEFT~\cite{Haidenbauer:2017dua} do not agree with the lattice QCD data at low energies. 

In this work, we revisit the fits to the lattice QCD data and the corresponding extrapolations to the physical point. We study in detail the differences between the non-relativistic ChEFT and covariant ChEFT in the description of the $\Lambda_c N$ $^3D_1$ phase shifts and inelasticity $\eta_1$, including the effects of baryon masses and $SD$ coupling in the contact terms, and the retardation effects in the one meson exchange term.~\footnote{We found that the retardation effects are quite small and therefore refrain from explicit discussions about these effects from now on, but they are always included in our study.} In addition, we study extrapolations to the physical point employing different fitting strategies to the lattice QCD data. These results are important to better understand the $\Lambda_cN$ interaction and might be helpful to guide future hypernuclei  experiments.

The paper is organized as follows. In Sec.~II, we briefly introduce the non-relativistic and the covariant chiral EFT. In Sec.~III we perform fits to the lattice QCD data of Ref.~\cite{Miyamoto:2019mfk}, focusing on the low energy region, where ChEFT is expected to  work. We summarize in Sec.~IV. 

\section{theoretical framework}
In this section, we briefly introduce the  non-relativistic ChEFT and covariant ChEFT for the $Y_cN$ interactions, where $Y_c=\Lambda_c,\Sigma_c$, and highlight the differences relevant for the present study. 

In the non-relativistic ChEFT, the next-to-leading order potentials consist of non-derivative four-baryon contact terms (CT) and one-meson-exchanges (OME).
The CT potentials for the $^1S_0$ and $^3S_1-{}^3D_1$ partial waves are~\cite{Haidenbauer:2020kwo}
\begin{align}\label{VCT_NONREL}
&V_{\textrm{CT},1S0}^{Y_cN}  = \tilde{C}_{^1S_0}+\tilde{D}_{^1S_0}m_{\pi}^2+(C_{^1S_0}+D_{^1S_0}m_{\pi}^2)(p^2+p'^2),\nonumber\\
&V_{\textrm{CT},3S1}^{Y_cN}  = \tilde{C}_{^3S_1}+\tilde{D}_{^3S_1}m_{\pi}^2+(C_{^3S_1}+D_{^3S_1}m_{\pi}^2)(p^2+p'^2),\nonumber\\
&V_{\textrm{CT},3D1-3S1}^{Y_cN}  = C_{\varepsilon_1}p'^2, \nonumber\\
&V_{\textrm{CT},3S1-3D1}^{Y_cN}  = C_{\varepsilon_1}p^2,
\end{align}
where  $p = |\bm{p}|$ and $p' = |\bm{p'}|$ are the initial and final center-of-mass (c.m.) momenta of the $Y_{c}N$ system, respectively. $\tilde{C}_{i}$, $\tilde{D}_{i}$, $C_{i}$, $D_{i}$($i=~^1S_0$, $^3S_1$, and $\varepsilon_1$) are LECs that need to be fixed by fitting to either experimental or lattice QCD data. The OME potential reads,
\begin{align}\label{VOMES_NONREL}
V_{\rm{OME}}^{Y_cN\rightarrow Y_c'N} =-\frac{g^{Y_cY_c'}_{A}g^{NN}_{A}}{4f^2_{\pi}} \frac{(\bm{\sigma_1}\cdot \bm{q})(\bm{\sigma_2}\cdot \bm{q})}{\bm{q}^2 + m_{\pi}^2} \times \mathcal{I}_{Y_cN\rightarrow Y_c'N},
\end{align}
where $\bm{q} = \bm{p}^{\prime}-\bm{p}$ is the transferred momentum. The coupling constants $g^{Y_cY_c'}_{A}$ and $g^{NN}_{A}$ and the isospin factor $\mathcal{I}$ can be found in, e.g., Refs.~\cite{Polinder:2006zh,Haidenbauer:2017dua}. The scattering amplitudes are then obtained by solving the coupled-channel Lippmann-Schwinger equation, 
\begin{align}\label{LS}
T_{\rho'\rho}^{\nu'\nu,J}(p',p;\sqrt{s})
=   V_{\rho'\rho}^{\nu'\nu,J}(p',p) +
  \sum_{\rho'',\nu''}\int_0^\infty \frac{dp''p''^2}{(2\pi)^3}  V_{\rho'\rho''}^{\nu'\nu'',J}(p',p'')\times
   \frac{2\mu_{\rho''}}{p_{\rho}^2-p''^2+i\eta}T_{\rho''\rho}^{\nu''\nu,J}(p'',p;\sqrt{s}),
\end{align}
where the labels $\nu,\nu',\nu''$ denote the particle channels,  $\rho,\rho',\rho''$ denote the partial waves, and $\mu_\rho$ is the pertinent reduced mass.
The on-shell momentum in the intermediate state, $p_\rho$, is
defined by $\sqrt{s}=\sqrt{M^{2}_{B_{1,\rho}}+p_{\rho}^{2}}+\sqrt{M^{2}_{B_{2,\rho}}+p_{\rho}^{2}}$. The potentials are regularized with an exponential form factor,
\begin{align}\label{EF}
  f_{\Lambda_F}(p,p') = \exp \left[-\left(\frac{p}{\Lambda_F}\right)^{4}-\left(\frac{p'}{\Lambda_F}\right)^{4}\right],
\end{align}
where $\Lambda_F$ is the cutoff whose value is in the range of $500-600$ MeV. 
The partial wave $S$ matrix is related to the on-shell $T$ matrix by

\begin{equation}
  S_{\rho'\rho}^{\nu'\nu} = \delta_{\rho'\rho} \delta^{\nu'\nu} - 2 i a T_{\rho'\rho}^{\nu'\nu},\quad a= \frac{{\sqrt{\bm{p}_\mathrm{cm}^{\nu'} \bm{p}_\mathrm{cm}^\nu} \mu^{\nu' \nu}}}{16\pi^2},
\end{equation}
where $\bm{p}_\mathrm{cm}$ is the C.M. three-momentum of the $\Lambda_c N$ system. The phase space factor $a$ is determined by the elastic unitarity of the scattering equation.
For single channels, the phase shifts $\delta$ can be obtained from the on-shell $S$ matrix,
\begin{equation}
  S = \exp(2i\delta).
\end{equation}

In order to calculate the phase shifts in coupled channels ($J>0$), we use the ``Stapp''- or ``bar''- phase shifts parametrisation~\cite{Stapp:1956mz} of the $S$ matrix, which can be written as
\begin{eqnarray}
  S &=& \left(
       \begin{array}{cc}
         S_{--} & S_{-+}\\
         S_{+-} & S_{++} \\
       \end{array}
     \right)\nonumber\\
   &=& \left(
       \begin{array}{cc}
         \exp(i\delta_-) & 0 \\
         0 & \exp(i\delta_+) \\
       \end{array}
     \right)
  \left(
       \begin{array}{cc}
         \cos(2\epsilon)  & i\sin(2\epsilon) \\
         i\sin(2\epsilon) & \cos(2\epsilon) \\
       \end{array}
     \right)
     \left(
       \begin{array}{cc}
         \exp(i\delta_-) & 0 \\
         0 & \exp(i\delta_+) \\
       \end{array}
     \right),
\end{eqnarray}
where the subscript ``$+$'' is $J+1$, ``$-$'' for $J-1$. The resulting phase shifts and mixing angles are
\begin{equation}
  \tan(2\delta_{\pm}) = \frac{\mathrm{Im}(S_{\pm\pm}/\cos(2\epsilon_J))} {\mathrm{Re}(S_{\pm\pm}/\cos(2\epsilon_J))}, \quad
  \tan(2\epsilon_J) = \frac{-iS_{+-}}{\sqrt{S_{++}S_{--}}}.
\end{equation}

For more details about the non-relativistic ChEFT, please refer to Refs.~\cite{Polinder:2006zh,Polinder:2007mp,Haidenbauer:2005zh,Haidenbauer:2007ra,Haidenbauer:2009qn,Haidenbauer:2013aj,Haidenbauer:2013oca,Haidenbauer:2015zqb, Haidenbauer:2017dua,Haidenbauer:2017sws,Haidenbauer:2019boi,Haidenbauer:2020kwo,Haidenbauer:2020uci,Haidenbauer:2021tlk}.

In the covariant ChEFT, as discussed in Ref.~\cite{Song:2020isu}, the $^1S_0$ and $^3S_1-{}^3D_1$ CT potentials for the $Y_cN$ system read,
\begin{align}\label{VCT_REL}
V_{\textrm{CT},1S0}^{Y_cN} & = \xi _{{Y_c N}} \left[C_{{1S0}} \left(R_{p'}^N R_{p'}^{{Y_c}}+R_p^N R_p^{{Y_c}}\right)+C'_{{1S0}} \left(R_{p'}^N R_p^N R_{p'}^{{Y_c}} R_p^{{Y_c}}+1\right)\right],\nonumber\\
V_{\textrm{CT},3S1}^{Y_cN} & = \frac{1}{9} \xi _{{Y_cN}} \left\{2\left(C_{{1S0}}- C'_{{1S0}} \right)\left( R_{p'}^{{Y_c}} R_p^{{Y_c}}- R_{p'}^N R_p^N\right)  \right.\nonumber\\
& \quad \left.+C_{{3S1}} \left(-6 R_{p'}^N R_p^N+9 R_{p'}^N R_{p'}^{{Y_c}}+9 R_p^N R_p^{{Y_c}}+6 R_{p'}^{{Y_c}} R_p^{{Y_c}}\right)\right.\nonumber\\
& \quad \left.+9 C'_{{3S1}} \left[R_{p'}^{{Y_c}} R_p^{{Y_c}} \left(R_{p'}^N R_p^N-2\right)+2 R_{p'}^N R_p^N+9\right]\right\},\nonumber\\
V_{\textrm{CT},3D1-3S1}^{Y_c N} & = \frac{\xi _{{Y_c N}}}{9 \sqrt{2}} \left\{\left(C_{{1S0}} - C'_{{1S0}}\right) \left[R_p^N \left(R_{p'}^N+3 R_{p'}^{{Y_c}}\right)-R_p^{{Y_c}} \left(3 R_{p'}^N+R_{p'}^{{Y_c}}\right)\right]\right.\nonumber\\
& \quad  \left.+C_{{3S1}} \left[9 R_p^{{Y_c}} \left(R_{p'}^N+4 R_p^N\right)+3 R_{p'}^N R_p^N-3 R_{p'}^{{Y_c}} \left(3 R_p^N+R_p^{{Y_c}}\right)\right]\right.\nonumber\\
&\quad \left.+9 C'_{{3S1}} \left\{R_{p'}^{{Y_c}} \left[R_p^N \left(4 R_{p'}^N R_p^{{Y_c}}+3\right)+R_p^{{Y_c}}\right]-R_{p'}^N \left(R_p^N+3 R_p^{{Y_c}}\right)\right\}\right\},\nonumber\\
V_{\textrm{CT},3S1-3D1}^{Y_cN} & = \frac{\xi _{{Y_cN}}}{9 \sqrt{2}} \left\{\left(C_{{1S0}}-C'_{{1S0}}\right) \left[R_{p'}^N \left(R_{p}^N+3 R_{p}^{{Y_c}}\right)-R_{p'}^{{Y_c}} \left(3 R_{p}^N+R_{p}^{{Y_c}}\right)\right]\right.\nonumber\\
& \quad  \left.+C_{{3S1}} \left[9 R_{p'}^{{Y_c}} \left(R_{p}^N+4 R_{p'}^N\right)+3 R_{p}^N R_{p'}^N-3 R_{p}^{{Y_c}} \left(3 R_{p'}^N+R_{p'}^{{Y_c}}\right)\right]\right.\nonumber\\
&\quad \left.+9 C'_{{3S1}} \left\{R_{p}^{{Y_c}} \left[R_{p'}^N \left(4 R_{p}^N R_{p'}^{{Y_c}}+3\right)+R_{p'}^{{Y_c}}\right]-R_{p}^N \left(R_{p'}^N+3 R_{p'}^{{Y_c}}\right)\right\}\right\},\nonumber\\
V_{\textrm{CT},3D1}^{Y_cN} & = \frac{2}{9} \xi _{{Y_cN}} \left\{\left(C_{{1S0}}-C'_{{1S0}}+3 C_{{3S1}}\right) \left(R_{p'}^N R_p^N-R_{p'}^{{Y_c}} R_p^{{Y_c}}\right) \right.\nonumber\\
& \quad \left.+9 C'_{{3S1}} \left[R_{p'}^N R_p^N \left(4 R_{p'}^{{Y_c}} R_p^{{Y_c}}-1\right)+R_{p'}^{{Y_c}} R_p^{{Y_c}}\right]\right\},
\end{align}
where
\begin{align*}
{\xi_{Y_cN}} = 4 \pi \frac{\sqrt{\left(E_{p'}^{Y_c}+M_{Y_c}\right)\left(E_{p}^{Y_c}+M_{Y_c}\right)\left(E_{p'}^{N}+M_{N}\right)\left(E_{p}^{N}+M_{N}\right)}}{4M_N M_{Y_c}}~~~~
\textrm{and}
~~~~R_{p(p')}^{Y_c,N}=\frac{p(p')}{E_{p(p')}^{Y_c,N}+M_{Y_c,N}}.
 \end{align*}
It should be noted that there exist three differences between the covariant and the non-relativistic ChEFT potentials as presented above: (1) the covariant chiral potentials explicitly contain the baryon masses $M_{Y_c}$ and $M_N$, where $M_{Y_c}=(M_{\Lambda_c}+M_{\Sigma_c})/2$, whose values are the same as those given in Table I of Ref.~\cite{Song:2020isu}; (2) because of the fact $M_{Y_c} \neq M_N$, the LECs from the $^1S_0$ partial wave also contribute to those of the $^3S_1-{}^3D_1$ partial waves; (3) the LECs responsible for the $SD$ coupling are correlated with those of the $^1S_0$ and $^3S_1$ potentials. It should be noted that the contributions from the  $\Sigma_cN$ intermediate state in the CT potentials were set  zero in both the non relativistic ChEFT~\cite{Haidenbauer:2017dua} and covariant ChEFT~\cite{Song:2020isu}, since the limited lattice QCD data could not fix these contributions.

The leading-order OME potential reads,
\begin{align}\label{VOMES_REL}
V_{\rm{OME}}^{Y_cN\rightarrow Y_c'N} =&-ig_{A}^{Y_c Y_c'} g_{A}^{NN}\bar{u}_{Y'_c}\left(p'\right)\left(\frac{\gamma^{\mu} \gamma_5 q_{\mu}}{2f_{\pi}}\right)u _{Y_c}(p) \frac{i}{\Delta E^2 - q^2 - m^2+i\epsilon}\nonumber\\ 
~&\times\bar{u}_N\left(-p'\right) \left(\frac{ \gamma^{\nu} \gamma_5 \boldmath{q}_{\nu}}{2f_{\pi}}\right) u_N(-p)  \times \mathcal{I}_{Y_cN\rightarrow Y_c'N},
\end{align}
where $\Delta E= E_{p'}-E_{p}$ is the transferred kinetic energy, i.e., the retardation effect, and we adopt the complete form of the Dirac spinor for the baryons involved
\begin{align*}
u_{B}(\bm{p},s)=
\left(
  \begin{array}{ccc}
    1 \\
    \frac{\bm{\sigma} \cdot \bm{p}}{E_p+M_B}\\
  \end{array}
\right)\chi_s.
\end{align*}
The coupled-channel Kadyshevsky equation~\cite{Kadyshevsky:1967rs} is solved to obtain the scattering amplitudes,
\begin{align}\label{SEK}
  T_{\rho\rho'}^{\nu\nu',J}(p',p;\sqrt{s})
  =&
   V_{\rho\rho'}^{\nu\nu',J}(p',p)+
  \sum_{\rho'',\nu''}\int_0^\infty \frac{dp''p''^2}{(2\pi)^3} \frac{M_{B_{1,\nu''}}M_{B_{2,\nu''}}~ V_{\rho\rho''}^{\nu\nu'',J}(p',p'')~
   T_{\rho''\rho'}^{\nu''\nu',J}(p'',p;\sqrt{s})}{E_{1,\nu''}E_{2,\nu''}
  \left(\sqrt{s}-E_{1,\nu''}-E_{2,\nu''}+i\epsilon\right)},
\end{align}
where $\sqrt{s}$ is the total energy of the two-baryon system in the center-of-mass frame and $E_{n,\nu''}=\sqrt{p''^{2}+M^{2}_{B_{n,\nu''}}}$, $(n=1,2)$. In the numerical study, the potentials are regularized with the same exponential form factor as that of  Eq.~(\ref{EF}). The relation between the phase shifts and $T$-matrix is the same as explained above except for the phase space factor $a$, which appearing in the Kadyshevsky equation is $a = \frac{1}{8\pi^2}\frac{\sqrt{\bm{p}_\mathrm{cm}^\nu \bm{p}_\mathrm{cm}^{\nu'} M_{B_1,\nu'} M_{B_2,\nu'} M_{B_1,\nu} M_{B_2,\nu}}}{(E_{1,\nu'}+E_{2,\nu'})(E_{1,\nu}+E_{2,\nu})}$ . More details about the covariant ChEFT approach can be found in Refs.~\cite{Ren:2016jna,Li:2016mln,Ren:2017yvw,Song:2018qqm,Ren:2018xxd,Li:2018tbt,Wang:2020myr,Bai:2020yml,Liu:2020uxi,Song:2021yab,Liu:2022nec,Lu:2021gsb}.

\section{fitting procedure}

In Ref.~\cite{Miyamoto:2017ynx}, the HAL QCD Collaboration presented the $^1S_0$ and $^3S_1$ phase shifts of the $\Lambda_c N$ interaction obtained from lattice QCD simulations with  $m_\pi=410, 570$, and $700$ MeV. In addition, the corresponding $^3D_1$ partial wave phase shifts and inelasticity $\eta_1$ can be found in the Ph.D. thesis of Takaya Miyamoto~\cite{Miyamoto:2019mfk}. The results show that the $\Lambda_cN~^3D_1$ phase shifts for $M_{\pi}=410$, $570$, and $700$~MeV are slightly repulsive for the center of mass energy no larger than $15$, $30$, and $40$~MeV, respectively and become attractive as $E_{\textrm{c.m.}}$ increases, and the inelasticity $\eta_1$ is close to unity in the whole energy region. Fittings to the $S$ partial waves of $m_\pi=410$ and $570$ MeV (with $E_{\textrm{c.m.}}\leq 30$ MeV), and extrapolations to the physical point were performed in  both the non-relativistic ChEFT~\cite{Haidenbauer:2017dua} and covariant ChEFT~\cite{Song:2020isu}. The predictions for the $^3D_1$ phase shifts and inelasticity $\eta_1$ turn out to be dramatically different. The $^3D_1$ interaction in the former approach is attractive, while that in the latter is repulsive. In addition,  both approaches predict a $SD$ coupling stronger than that shown by the lattice QCD data.

In this study, we first investigate where such differences in the predicted $\Lambda_cN$ $^3D_1$ phase shifts between the two approaches originate. In particular, we focus on the masses of $Y_c$ and $N$ and the $SD$ coupling in the CT potential. We note that there are no baryon mass terms in the CT potential of the non-relativistic ChEFT, while $M_B$ ($M_{Y_c}$, $M_N$) appears in the baryon spinors of the covariant ChEFT. As $M_{Y_c}\ge M_N$, we used the ``physical'' masses for $Y_c$ and $N$ in our previous study, which has the consequence that $C_{1S0}$ ($C'_{1S0}$) also contributes to the $^3S_1-{}^3D_1$ partial waves~\cite{Song:2020isu}. In addition, the $SD$ coupling in the covariant ChEFT is correlated to the $^3S_1$ potential, while a free LEC appears in the non-relativistic ChEFT. These two differences lead to in total $2^2=4$ combinations that will be examined. In addition to our previous study~\cite{Song:2020isu}, we perform three more fits to the same lattice QCD data,  and make a systematic comparison of the results, to better understand how the results depend on the baryon masses and $SD$ coupling in the CT potential.

Moreover, since both approaches fail to precisely reproduce the lattice QCD $^3D_1$ phase shifts of $\Lambda_cN$ at low energies, we adopt a new fitting strategy where the phase shifts of $\Lambda_cN$ $^3S_1$, $^3D_1$ partial waves and inelasticity $\eta_1$ with $E_{\textrm{c.m.}}\leq 5$ MeV are simultaneously fitted.
The new strategy can provide a closer look at the two approaches in the descriptions of low energy lattice QCD data. Note that we only consider the lattice QCD data with $m_\pi=410$ and $570$ MeV in all the aforementioned fittings. Details of the fitting strategies in this work are shown in Table~\ref{fitting_conditions}.
\begin{table}[H]
\centering
\caption{Seven fitting strategies studied in this work, where $\checkmark$ indicates that the $SD$ coupling in the CT potential is turned on, while $\times$ denotes that the $SD$ coupling is turned off in the covariant ChEFT approach.}\label{fitting_conditions}
\setlength{\tabcolsep}{12pt}
\begin{tabular}{m{0.8cm}<{\centering}m{3.2cm}<{\centering}m{2.2cm}<{\centering}m{1.8cm}<{\centering}m{1.8cm}<{\centering}}
\hline
\hline
Strategy & Lattice QCD data fitted & Approach &  $M_{Y_c}[~~~~]M_{N}$ & $SD$ coupling \\
\hline
$1$ & \multirow{3}{*}{$^3S_1$} & \multirow{4}{*}{cov. ChEFT} &  $\neq$ & $\checkmark$  \\
$2$ &  &  &$\neq$ & $\times$  \\
$3$ & \multirow{1}{*}{$E_\textrm{{c.m.}}\leq30$~MeV} & &$=$ & $\checkmark$ \\
$4$ &  &  &$=$ & $\times$  \\
\cline{1-5}$5$ & \multirow{2}{*}{$^3S_1$, $^3D_1$, $\eta_1$} & \multirow{2}{*}{cov. ChEFT} &   $\neq$ & $\checkmark$  \\
$6$ & \multirow{2}{*}{$E_\textrm{{c.m.}}\leq5$~MeV}  &  &   $=$ & $\checkmark$  \\
$7$ &       & non-rel. ChEFT &    &  \\
\hline
\hline
\end{tabular}
\end{table}

\section{Results and discussions}

\subsection{Origin of the difference in predicting the $\Lambda_cN$ $^3D_1$ phase shifts}

\begin{figure}[H]
  \centering
  \includegraphics[width=0.8\textwidth]{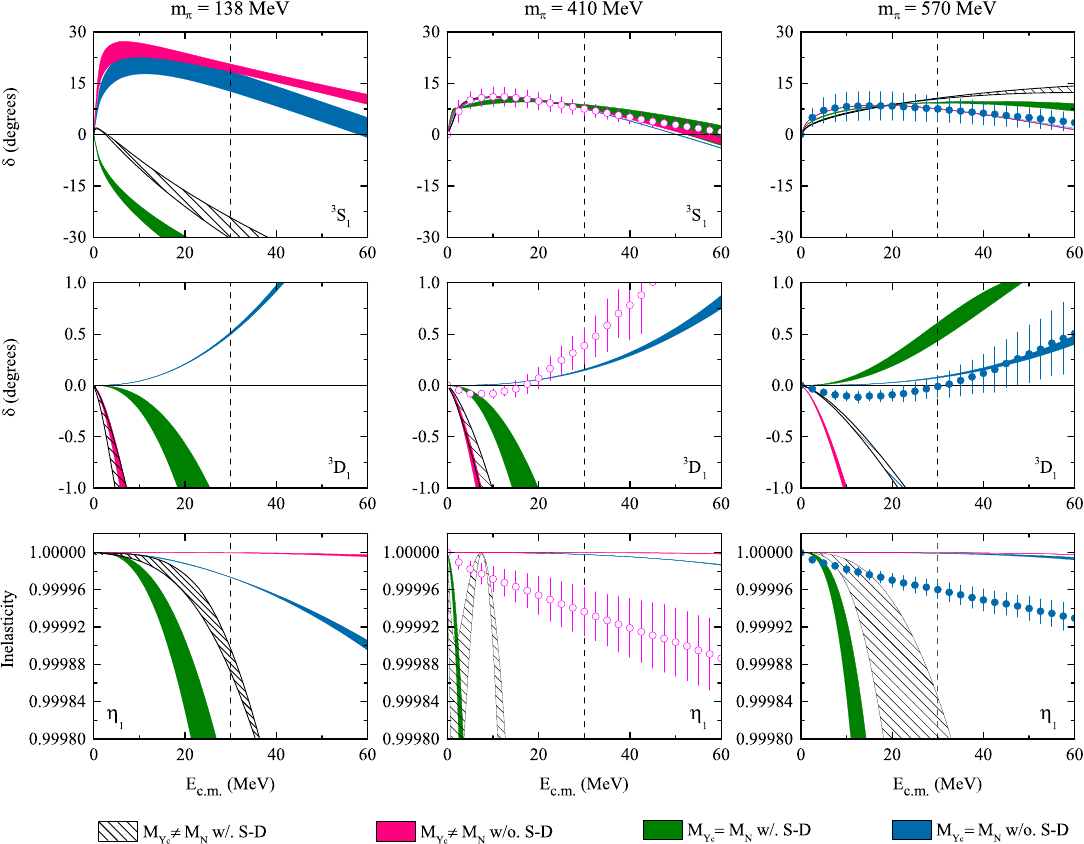}
  \caption{$\Lambda_c N$ $^3S_1$, $^3D_1$ phase shifts and inelasticity $\eta_1$ for different pion masses. The results are obtained by fitting to the lattice QCD $\Lambda_c N$ $S-$wave phase shifts for $E_{\mathrm{c.m.}}\leq 30$ MeV. The bands are generated from the variation of $\Lambda_F$ from $600$ MeV to $700$ MeV. Different labels denote the $\Lambda_cN$ phase shifts of strategies $1-4$: ``w/.'' is the abbreviation for ``with'', and ``w/o.'' is the abbreviation for ``without''.}\label{only_fit_3s1_Fig}
\end{figure}

The fitted results of strategies $1-4$ as described in the previous section are summarized qualitatively in Table~\ref{only_fit_3s1_Table} and quantitatively in Fig.~\ref{only_fit_3s1_Fig}. It is noted that the treatment of the potentials in strategy $1$ is that adopted in Ref.~\cite{Song:2020isu}~\footnote{The $\chi^2$ shown in Table~\ref{only_fit_3s1_Table} is larger than that in  Ref.~\cite{Song:2020isu} because of different fitting strategies.  The $\chi^2$ in  Ref.~\cite{Song:2020isu} is obtained by fitting to the $^3S_1$ 
phase shits, while the $\chi^2$ in Table~\ref{only_fit_3s1_Table} includes the $^3D_1$ and mixing angle data as well.}, and strategy $4$ is approximately the same as that of the non-relativistic ChEFT. The following conclusions can be obtained from the table: first, the baryon masses affect the $^3D_1$ phase shifts for the large pion mass ($m_\pi=570$ MeV), where negative phase shifts are obtained in strategies $1$, $2$ and they become positive if $M_{Y_c}$ is taken to be the same as $M_{N}$ (strategies $3$, $4$). Second, only when $M_{Y_c}=M_{N}$ and the $SD$ coupling in the CT potential is turned off, the $^3D_1$ interaction becomes attractive in the unphysical region (strategy $4$). Third, the $SD$ coupling in the covariant ChEFT reduces the attraction in the $^3S_1$ partial wave in the physical region, compared with the non-relativistic case, as shown in strategies $1$ and $3$.

\begin{table}[H]
\centering
\caption{Dependence of the  $\Lambda_c N~^3S_1$ and $~^3D_1$ phase shifts  on the baryon masses and $SD$ coupling for different pion masses (in units of MeV). The ``$+$'' and ``$-$'' indicate  the  sign of the $\Lambda_c N~^3S_1$ and $~^3D_1$ phase shifts within the fitting region $E_{\mathrm{c.m.}}\leq 30$~MeV, where  ``$+$'' and ``$-$'' denote attractive and repulsive potentials, respectively. The values of the $\chi^2/\textrm{d.o.f.}$ (in units of $10^{-2}$) are obtained with $\Lambda_{F} = 600/700$ MeV.}\label{only_fit_3s1_Table}
\setlength{\tabcolsep}{10pt}
\begin{threeparttable} 
\begin{tabular}{m{0.8cm}<{\centering}m{1.8cm}<{\centering}m{1.8cm}<{\centering}m{0.8cm}<{\centering}m{0.4cm}<{\centering}m{0.4cm}<{\centering}m{1.2cm}<{\centering}}
\hline
\hline
Strategy & $M_{Y_c}[~~~~]M_{N}$ & $SD$ coupling & $m_\pi$ & $\delta_{^3S_1}$ & $\delta_{^3D_1}$ & $\chi^2/\textrm{d.o.f.}^{\ddag}$\\
\hline
\multirow{3}{*}{$1$} & \multirow{3}{*}{$\neq$} & \multirow{3}{*}{$\checkmark$} & $138$ & $+-^{\dag}$ & $-$  &  \\
                     &   &                               & $410$ & $+$ & $-$ & $1.30/1.32$  \\
                     &   &                               & $570$ & $+$ & $-$ & $25.9/30.9$ \\
\cline{3-7}\multirow{3}{*}{$2$} & \multirow{3}{*}{$\neq$} & \multirow{3}{*}{$\times$} & $138$ & $+$ & $-$  &  \\
                     &  &                            & $410$ & $+$ & $-$ & $1.30/1.08$ \\
                     &  &                            & $570$ & $+$ & $-$ & $0.16/0.08$ \\
\cline{2-7}\multirow{3}{*}{$3$} & \multirow{3}{*}{$=$} & \multirow{3}{*}{$\checkmark$} & $138$ & $-$ & $-$  &\\
                     &  &                             & $410$ & $+$ & $-$ & $3.41/18.8$ \\
                     &  &                             & $570$ & $+$ & $+$ & $7.31/11.0$ \\
\cline{3-7}\multirow{3}{*}{$4$} & \multirow{3}{*}{$=$} & \multirow{3}{*}{$\times$} & $138$ & $+$ & $+$ &  \\
                     &  &                        & $410$ & $+$ & $+$ & $2.41/2.07$ \\
                     &  &                        & $570$ & $+$ & $+$ & $0.17/0.14$ \\
\hline
\hline
\end{tabular}
      \begin{tablenotes}
		\item $^{\dag}$ indicate that the $\Lambda_cN~^3S_1$ interaction for $m_{\pi}=138$~MeV is weakly attractive only at the very low energy region (about $E_\mathrm{c.m.}=3$~MeV) and then becomes repulsive as the kinetic energy increases.
		\item $^{\ddag}$ The small $\chi^2$'s compared with those of Table~\ref{simultaneously_fit} imply that it is easy to reproduce the lattice QCD $^3S_1$ phase shifts than the coupled channel results.		
     \end{tablenotes}
\end{threeparttable}
\end{table}

\subsection{Simultaneous fits to the $\Lambda_c N$ $^3S_1-{}^3D_1$ partial waves}

In this subsection, we  simultaneously fit to the phase shifts of $\Lambda_cN$ $^3S_1$, $^3D_1$ and inelasticity $\eta_1$ of the lattice QCD data for  $m_\pi=410$ and $570$ MeV  with a smaller energy range from threshold up to $E_\mathrm{c.m.}=5$~MeV in order to achieve a $\chi^2/\mathrm{d.o.f.}\approx1$. With this new strategy, we  aim to check whether the covariant ChEFT approach or the non-relativistic ChEFT approach can precisely describe the lattice QCD data at low energies, where they are believed to work the best. In the covariant ChEFT, we only 
consider two strategies: either $M_{Y_c}\neq M_{N}$ or $M_{Y_c}=M_{N}$. The $SD$ coupling appears naturally  in the CT potentials, therefore we did not manually turn it off. The non-relativistic ChEFT approach is also applied to perform the fits for comparison. The fitted results of strategies $5-7$, as described in Table~\ref{fitting_conditions}, are qualitatively shown in Table~\ref{simultaneously_fit} and quantitatively shown in Fig.~\ref{simultaneously_fit_rel} and Fig.~\ref{simultaneously_fit_nonrel}.

\begin{table}[H]
\centering
\caption{Phase shifts of $\Lambda_c N~^3S_1$ and $^3D_1$ partial waves for different pion masses based on the covariant ChEFT approach and non-relativistic ChEFT approach. The former depends on the baryon masses used, either physical or the lattice QCD $M_{Y_c}$ and $M_N$ or their average. These results are obtained by fitting to the phase shifts of $^3S_1$, $^3D_1$ and inelasticity $\eta_1$ simultaneously for $E_\mathrm{c.m.}\leq5$~MeV from lattice QCD simulations, and  $m_{\pi}$ is in units of MeV. The ``$+$'' and ``$-$'' indicate the sign of $\Lambda_c N~^3S_1$ and $^3D_1$ partial waves phase shifts within the fitting region, where ``$+$'' and ``$-$'' stand for attractive and repulsive potentials, respectively. The values of the $\chi^2/\textrm{d.o.f.}$ are obtained with $\Lambda_{F} = 600/700$ MeV in the covariant ChEFT and $\Lambda_{F} = 500/600$ MeV in the non-relativistic ChEFT.} \label{simultaneously_fit}
\setlength{\tabcolsep}{10pt}
\begin{tabular}{m{0.8cm}<{\centering}m{2.2cm}<{\centering}m{1.8cm}<{\centering}m{0.8cm}<{\centering}m{0.4cm}<{\centering}m{0.4cm}<{\centering}m{1.2cm}<{\centering}}
\hline
\hline
Strategy & Approach & $M_{Y_c}[~~~~]M_{N}$ & $m_\pi$ & $\delta_{^3S_1}$ & $\delta_{^3D_1}$ & $\chi^2/\textrm{d.o.f.}$\\
\hline
\multirow{3}{*}{$5$} & \multirow{6}{*}{cov. ChEFT} & \multirow{3}{*}{$\neq$}   & $138$ & $+$ & $-$ &  \\
                     & &                          & $410$ & $+$ & $-$ & $2.65/0.92$ \\
                     & &                          & $570$ & $+$ & $-$ & $3.32/3.25$ \\
\cline{3-7}\multirow{3}{*}{$6$}&  & \multirow{3}{*}{$=$}& $138$ & $+$ & $+$ &  \\
                     & &                        & $410$ & $+$ & $+$ & $3.29/3.30$ \\
                     & &                        & $570$ & $+$ & $+$ & $5.33/5.35$ \\
\hline
\multirow{3}{*}{$7$} & \multirow{3}{*}{non-rel. ChEFT} & \multirow{3}{*}{} & $138$ &  $-$ & $+$ &  \\
                    & &   & $410$ &  $+$ & $+$ & $2.22/2.25$ \\
                    & &   & $570$ &  $+$ & $+$ & $5.37/5.45$ \\
\hline
\hline
\end{tabular}
\end{table}

\subsubsection{Covariant ChEFT}

\begin{figure}[H]
  \centering
  \includegraphics[width=0.8\textwidth]{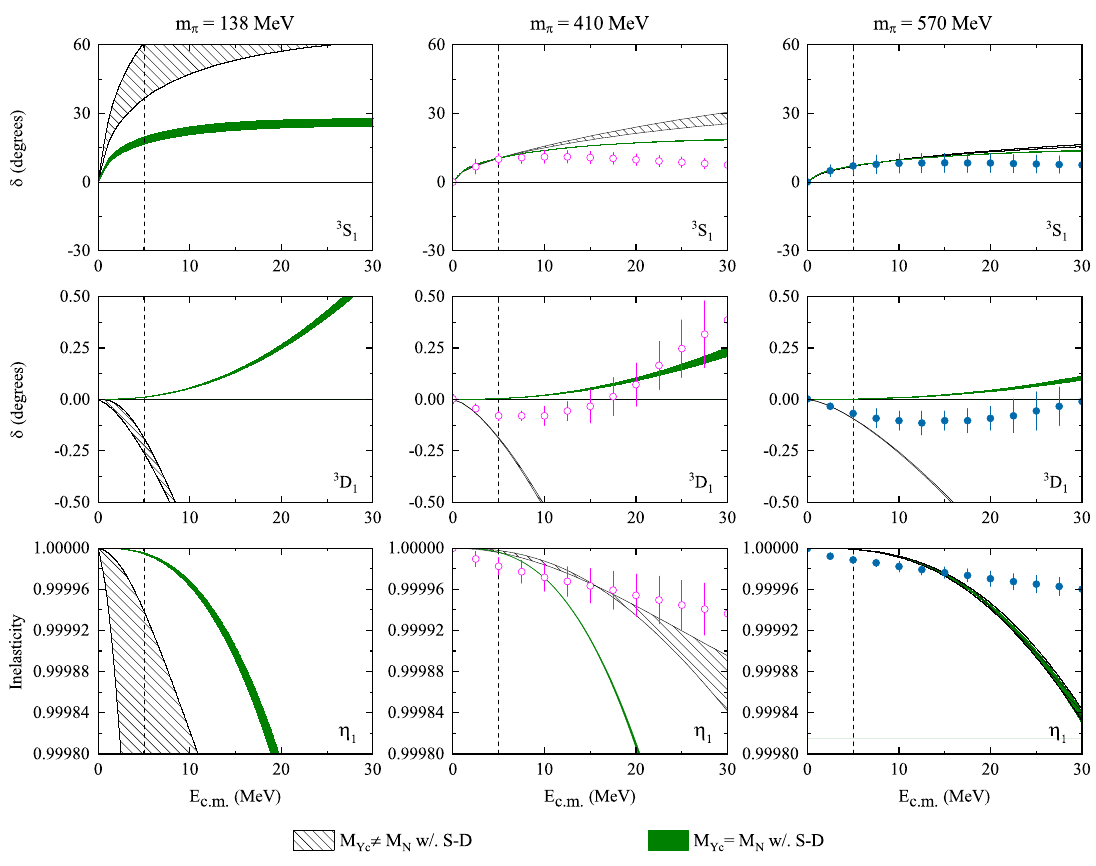}
  \caption{Same as Fig.~\ref{only_fit_3s1_Fig}, but the results are obtained by fitting to the lattice QCD phase shifts of $^3S_1$, $^3D_1$ partial waves and inelasticity $\eta_1$ simultaneously for $E_\mathrm{c.m.}\leq5$ MeV.}\label{simultaneously_fit_rel}
\end{figure}

First, we study how the use of ``physical" baryon masses   affects the description of the $\Lambda_cN$ interactions in the covariant ChEFT. The relevant fitting details and the corresponding values of the $\chi^2/\mathrm{d.o.f.}$ are summarized in Table~\ref{simultaneously_fit}. For strategy $5$, with lattice QCD  $M_{Y_c}, M_N$  in the $\Lambda_c N$ CT potentials within the fitting region $E_{\mathrm{c.m.}}\leq 5$~MeV, we presented the phase shifts of $\Lambda_c N~^3S_1$ and $^3D_1$ partial waves and inelasticity in Fig.~\ref{simultaneously_fit_rel}. One can see that the $\Lambda_c N$ $^3S_1$ and $^3D_1$ phase shifts  agree quantitatively with the lattice QCD data within uncertainties, and the asymptotic behaviors of inelasticity are in good agreement with the lattice QCD data. Comparing these results with those of strategy $6$ where $M_{Y_c}=M_N$, shown in Fig.~\ref{simultaneously_fit_rel}, one can see that the $\Lambda_cN~^3D_1$ interactions are attractive, contrary to the repulsive potential obtained in strategy 5.

In both cases, the extrapolation of the relativistic $\Lambda_c N~^3S_1$ and $^3D_1$ partial waves phase shifts and inelasticity to the physical point shows that the $\Lambda_c N$ interaction is attractive in the $^3S_1$ partial wave within the fitting region. Comparing the above results with strategy $1$ (our previous study), where the $\Lambda_cN~^3S_1$ potential is repulsive, we conclude that the extrapolated phase shifts of $\Lambda_cN~^3S_1$ are not very stable.

\begin{figure}[H]
  \centering
  \includegraphics[width=0.8\textwidth]{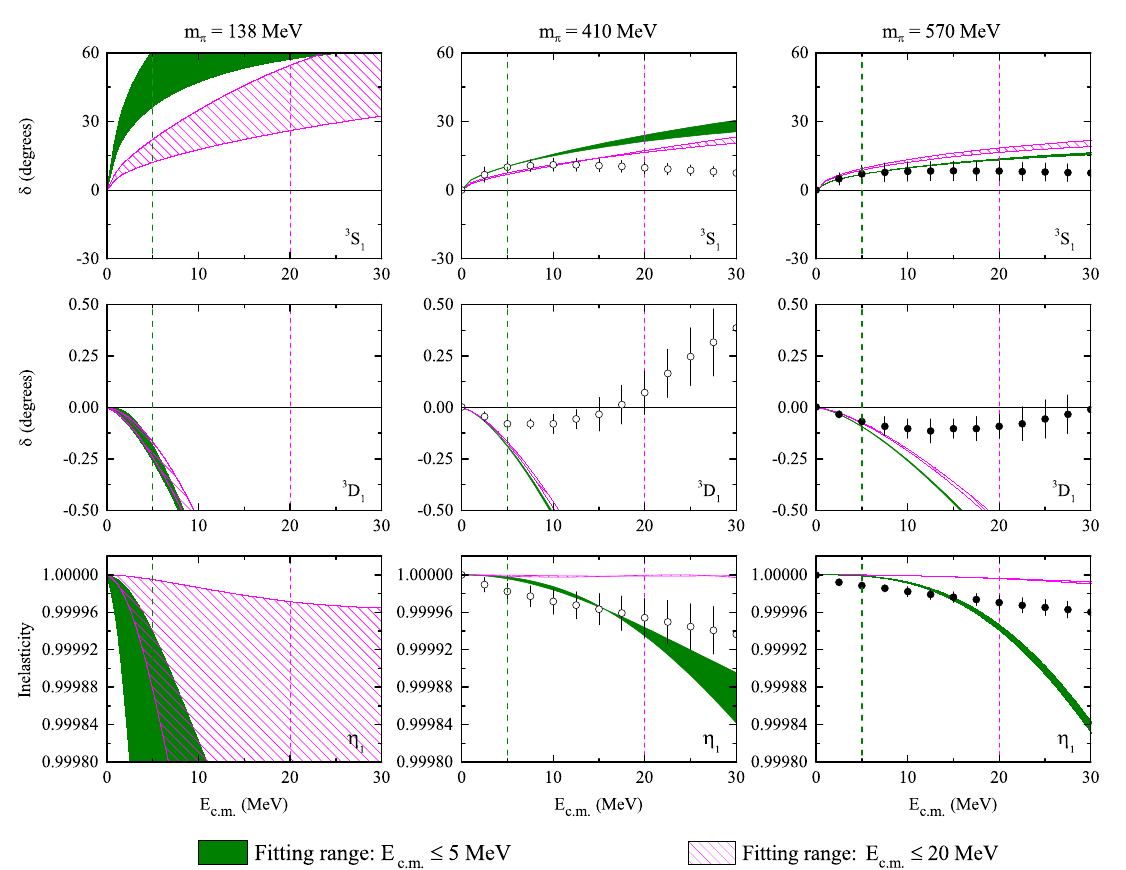}
  \caption{$\Lambda_c N$ $^3S_1$, $^3D_1$ phase shifts and inelasticity $\eta_1$ as functions of $E_\mathrm{c.m.}$. The results are obtained by fitting to the lattice QCD $\delta_{3S1}$, $\delta_{3D1}$  and  $\eta_1$ simultaneously up to  $E_\mathrm{c.m.}\leq5$ MeV (blue bands) and $E_\mathrm{c.m.}\leq20$ MeV (magenta bands) in covaraint ChEFT.}\label{simultaneously_fit_rel_20}
\end{figure}

To investigate whether the energy region fitted can affect the extrapolations, we also fitted the lattice QCD data up to  $E_\mathrm{c.m.}\le20$ MeV in the covariant ChEFT approach. The results are shown in Fig.~\ref{simultaneously_fit_rel_20} in comparison with the results obtained by fitting only up to $E_\mathrm{c.m.}\le 5$ MeV. The two fits are qualitatively consistent with each other . Only $\eta_1$ is closer to unity in the new fit. In addition, the extrapolated $\delta_{3S1}$ and $\eta_1$  show some visible differences. At $m_\pi=138$ MeV, $\delta_{3S1}$ becomes smaller, and $\eta_1$ becomes more dependent on the cutoff. 
\subsubsection{Non-relativistic ChEFT}
\begin{figure}[H]
\centering
\includegraphics[width=0.8\textwidth]{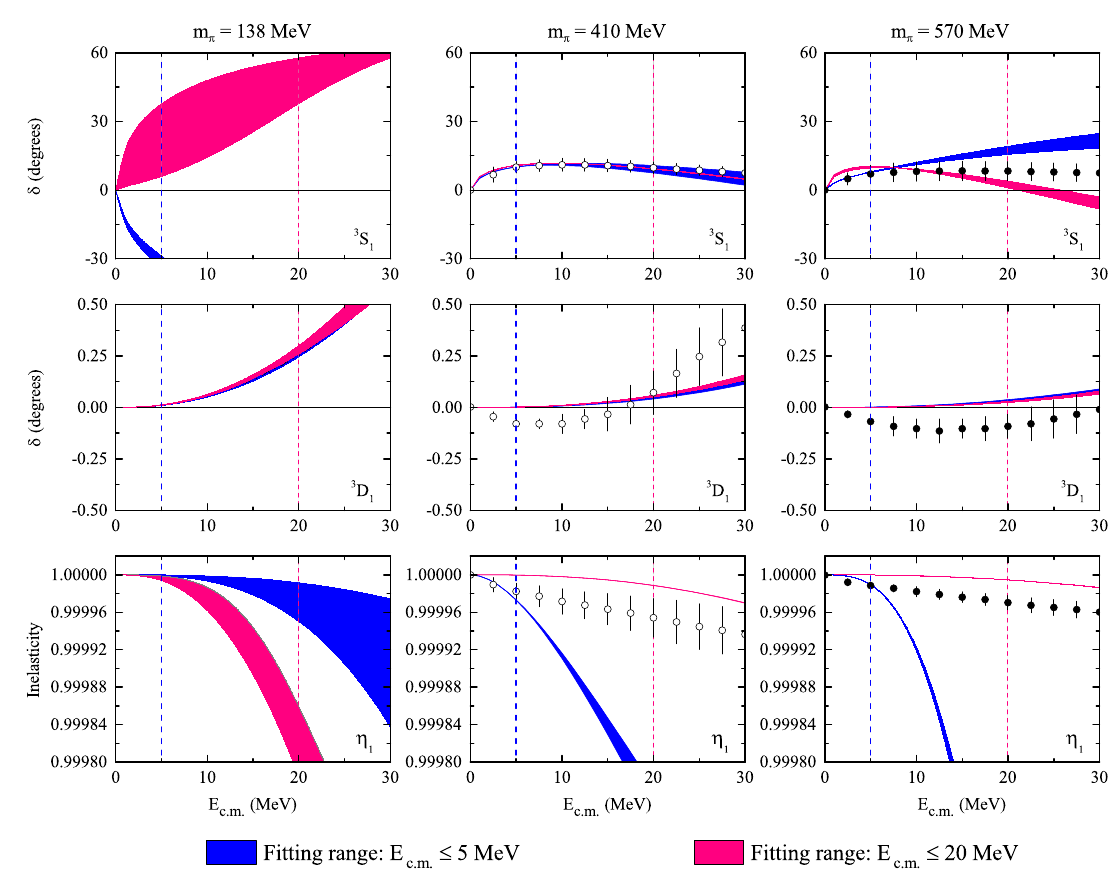}
\caption{Same as Fig.~\ref{simultaneously_fit_rel_20}, but for the  non-relativistic ChEFT.}
\label{simultaneously_fit_nonrel}
\end{figure}

Focusing on the lattice QCD data with $E_{\mathrm{c.m.}}\leq 5$~MeV, we show in Fig.~\ref{simultaneously_fit_nonrel} (the blue bands) the $\Lambda_cN~^3S_1$,$^3D_1$ partial wave phase shifts and inelasticity obtained from strategy $7$ in the non-relativistic ChEFT. The corresponding $\chi^2/\mathrm{d.o.f.}$ are listed in Table~\ref{simultaneously_fit}. Here, we find that the non-relativistic phase shifts of $\Lambda_c N~^3S_1$ partial wave and inelasticity are in qualitative agreement with the lattice QCD data in the region fitted. On the other hand,  the $\Lambda_cN~^3D_1$ phase shifts turn out to be positive, while the lattice QCD data are negative, though quite small. This is very different from the covariant case as shown in Fig.~3, where the $^3D_1$ phase shifts are negative for the energy region studied. According to the previous experience in the $NN$ sector~\cite{Bai:2020yml,Bai:2021uim},  the two EFTs should behave similarly in the low-energy regime, while the covariant EFT usually agrees better with the lattice QCD data than the non-relativistic EFT in the relatively high-energy regime. The present results are in conflict with such expectations to some extent. A better understanding can only be achieved once more precise lattice data with realistic uncertainties become available.  

When we extrapolate the non-relativistic results to the physical point, we find that the $\Lambda_c N$ interaction is repulsive in the $^3S_1$ partial wave within the fitting region. 
Comparing these results with those of the covariant ChEFT and the left panel in Fig.4 of Ref.~\cite{Haidenbauer:2017dua}, 
we again conclude that the extrapolated $\Lambda_cN~^3S_1$ interactions are not very stable, i.e., sensitive to the adopted fitting strategies.

The HB $\Lambda_c N$~$^3S_1$, $^3D_1$ phase shifts and inelasticity with LECs obtained by fitting to the lattice QCD data up to $E_\mathrm{c.m.}\le 20$ MeV (Fit 2) are compared to those obtained by fitting only up to $E_\mathrm{c.m.}\le5$ MeV (Fit 1)  in Fig.~\ref{simultaneously_fit_nonrel}. Compared to Fit 1, the descriptions of  $\delta_{3D1}$   remain almost unchanged in Fit 2, but the $^3S_1$ phase shifts are very different. In Fit 1, $\delta_{3S1}$ increases with $E_\mathrm{c.m.}$ for the case of $m_{\pi}=570$ MeV, while  in Fit 2,  it increases with $E_\mathrm{c.m.}$ for $E_\mathrm{c.m.}\leq 5$ MeV and then  decreases with $E_\mathrm{c.m.}$, which are in better agreement with the lattice QCD simulations at least for the energy region shown in this figure. Moreover, the difference for the case of $m_{\pi}=570$ MeV eventually contributes to the completely different prediction of the physical $\delta_{3S1}$, where the phase shifts become positive in Fit 2. As for the case of $m_{\pi}=410$ MeV, the two results show no qualitative difference. For the inelasticity, the results obtained in Fit 2 are  closer to unity  and larger than the lattice QCD simulations for the energy region  fitted  and become more  independent on  $E_\mathrm{c.m.}$ as the pion mass increases. 

\section{conclusion}

The $\Lambda_c N$ $^3S_1-{}^3D_1$ interactions were studied in leading order covariant ChEFT and next-to-leading order non-relativistic ChEFT. The low-energy constants were determined in two different strategies by fitting to the HAL  QCD lattice data, i.e. (a) by only fitting to the $\Lambda_c N$ $^3S_1$ partial wave phase shifts, (b) by a combined fit to the phase shifts of $^3S_1$, $^3D_1$ and inelasticty $\eta_1$. It was shown that for the first strategy, the predicted $\Lambda_c N$ $^3D_1$ phase shifts from the covariant ChEFT were consistent with the low-energy lattice QCD data  by using lattice QCD $M_{Y_c}$, $M_N$ and retaining the $SD$ coupling in the contact potentials, while for the second strategy, one obtained results similar to those of the first strategy in the covariant ChEFT. However,  the non-relativistic ChEFT predicts an attractive  $\Lambda_c N$ $^3D_1$ interaction in both cases, which is inconsistent with the low-energy lattice QCD data. In addition, we found that the extrapolated $\Lambda_cN$ $^3S_1$ phase shifts in the physical region were very sensitive to the fitting strategies and the theoretical approaches used. The covariant ChEFT predicts a repulsive/attractive $^3S_1$ interaction depending on the fitting strategy (a)/(b), while the non-relativistic ChEFT predicts the  opposite, which also depends on the energy region fitted. These results indicate that more refined lattice QCD data are needed to reach a firm conclusion about the $\Lambda_c N$ $^3S_1-{}^3D_1$ interactions.

It is necessary to point out that there are ongoing discussions on the validity of the HAL QCD method~\cite{Iritani:2017rlk,Beane:2017edf,Iritani:2018vfn}. In our present work, we have of course assumed that the method is valid and the $^3D_1$ phase shifts and particularly the inelasticity are correctly extracted with the precision claimed in Ref.~\cite{Miyamoto:2019mfk}.  Hopefully, the present study can motivate a closer look at the $\Lambda_c N$ interaction in the $^3S_1$-$^3D_1$ coupled channel. 
 
\section{Acknowledgements}

This work was partly supported by the National Natural Science Foundation of China (NSFC) under Grants No. 11975041, No.11735003, and No.11961141004. Yang Xiao acknowledges the support from China Scholarship Council.

\bibliographystyle{unsrt}
\bibliography{references.bib}
\end{document}